\documentclass[11pt,preprint]{aastex}

\received{}
\revised{}
\accepted{}

\shorttitle{Photometric Redshift Determination with the BATC Multicolor System}

\shortauthors{Xia et al.}

\begin{document}

\title{Photometric Redshift Determination with the BATC Multicolor System}

\author{Lifang Xia, Xu Zhou, Jun Ma, Hong Wu, Wei-Hsin Sun, 
Zhaoji Jiang, Suijian Xue, Jiansheng Chen, Wenping Chen}
\affil{National Astronomical Observatories,
Chinese Academy of Sciences, Beijing, 100012, P. R. China}
\email{xlf@vega.bac.pku.edu.cn}

\begin{abstract}

In this paper, we present the methodology of photometric redshift 
determination with the BATC 15-color system by using $hyperz$ 
program.
Both simulated galaxies and real galaxies with known redshifts were used
to estimate the accuracy of redshifts inferred from the multicolor 
photometry. 
From the test with simulated galaxies, the uncertainty in the inferred
redshifts is about $0.02\sim0.03$ for a given range of photometric
uncertainty of $0\hspace{0.1cm}{.}\hspace{-0.15cm}^{m}05 \sim
0\hspace{0.1cm}{.}\hspace{-0.15cm}^{m}10$.  The results 
with the 27 real galaxies are in good agreement with the simulated ones. 
The advantage of using BATC intermediate-band system to derive 
redshift is clear through the comparison with the 
$UBVRI$ broad-band system.  The accuracy in redshift
determination with BATC system is mainly affected by the selection of 
filters and the photometric uncertainties in the observation.  
When we take the limiting magnitudes of the 15 filters into account,
we find that redshift can be determined with good accuracy 
for galaxies with redshifts less than 0.5,
using only filters with central wavelengths shorter than 6000{\AA}.  
  
\end{abstract}


\section{INTRODUCTION}

In multicolor photometric surveys, the 
redshifts of a large number of objects in a given 
field can readily be obtained from the color information.  
Although multicolor photometry does not yield redshift information as 
accurately as spectroscopy does, it has the virtues of deeper limiting
magnitude, faster batch reduction, and better time-saving from the 
simultaneous determination of redshifts of many objects in a given field.  
With the redshifts determined for a large sample of galaxies via 
multicolor photometry, astronomers are able to study statistically 
the evolution of galaxies in number as well as in luminosity 
(Pascarelle et al. 1998; Volonteri et al. 2000;  Gal et
al. 2000).  In a simulation using 40 bands, the efficiency of 
photometric redshift determination 
for faint objects is comparable to slitless 
spectroscopy (Hickson et al. 1994). The techniques of photometric 
redshift are thus said to be not only the ``poor person's redshift machine'' 
but also the only viable way so far to acquire redshift information 
for a large quantity of faint objects, because the majority of these 
objects will still remain beyond the limit of spectroscopy in the foreseeable 
future (Bolzonella et al. 2000).

A number of computer codes performing photometry fitting have been
developed and applied to data acquired in several survey projects, 
such as HDF, SDSS, CADIS, etc (Sowards 
et al. 1999; Yahata et al. 2000; Wolf et al. 2001). 
Two methods have been widely used, one is the 
``Empirical Training Set'' method (Connolly et
al. 1995; Wang et al. 1999), the other is the 
``Spectral Energy Distribution'' (hereafter SED) fitting method . 

The empirical training set method determines redshifts by the
empirical linear relation between magnitudes (or colors)
and redshifts. Although this method requires no assumptions on
galaxy spectra and their evolution, {\bf there are still} 
a few shortages. For example, the
empirical relation changes with the data obtained with different
filter sets.  Furthermore, in high redshifts, the sample of 
spectroscopic templates becomes 
smaller and less complete, which make the redshift determination
less reliable.  The SED fitting method, on the other hand, 
is based on the fit of the
overall shape of a spectrum, i.e., it relies on the detection of
apparent spectral properties such as Lyman-forest and Balmer Jump, etc.
The fitting is performed by comparing the observed SEDs to the  
template spectra acquired using the same 
photometric system (Corbin et al. 2000; Fontana et al. 2000).

The BATC (Beijing-Arizona-Taipei-Connecticut) large-field sky 
survey in 15 intermediate-band colors commenced in 1994. 
Over the years, the survey has produced 
a database, which can be used to derive 
the redshifts of nearby galaxies between $z=0$ and $0.5$, providing
essential information regarding the structure of the local 
universe and the nearby galaxy clusters, especially Abell clusters
(Yuan et al. 2001). The purpose of the study in this paper is to estimate the
accuracy of $z_{phot}$ using the BATC 15-color photometric system.

The content of this paper is as follows. In $\S$ 2 we describe the BATC
photometric system and the observations of two fields used as the
real sample. The procedures of data reduction are given briefly in
$\S$ 3. The application of $z_{phot}$ code $hyperz$ is described in
$\S$ 4. In $\S$ 5, the comparison between 
the BATC system and the $\ UBVRI$ system using the simulation 
test is shown, along with the filter dependence
of $z_{phot}$'s. We compare the results of
$z_{phot}$'s with the spectroscopic redshifts $z_{spec}$'s in
$\S$ 6. Discussions and conclusions are presented in
$\S$ 7.

\section{THE BATC PHOTOMETRIC SYSTEM AND OBSERVATIONS} 

The BATC Sky Survey performs photometric observations with a large 
field multicolor system. 
The observation is carried out with the 60/90 cm f/3
Schmidt Telescope of National Astronomical Observatories, Chinese Academy
of Sciences, (NAOC)
located at the Xinglong station. A Ford Aerospace 2048$\times$2048 CCD
camera with 15$\micron$ pixel size is mounted at the main focus of the
Schmidt telescope. The field of view is 58$\times$58 $\rm
arcmin^{2}$ with a plate scale of 1.7 arcsec/pixel.

There are 15 intermediate-band filters in the BATC filter system, which
covers an optical wavelength range from 3000 to 10000{\AA}
(Fan et al. 1996; Zhou et al. 2001). The filters are specifically
designed to avoid contamination from most of the strong and 
variable night sky emission lines. The filter transmission curves 
are shown in Figure 1 and the corresponding parameters are tabulated
in Table 1.


As in the definition of the $\ AB_{\nu}$ system of Oke \& Gunn (1983), 
the magnitudes of the BATC system is defined as follows:
\begin{equation}
m_{\rm batc} = -2.5{\rm log}\widetilde{F_{\nu}} - 48.60,
\end{equation}
where  $\widetilde{F_{\nu}}$ is the flux per unit frequency in units of
$\rm erg~s^{-1}cm^{-2}Hz^{-1}$ (Fan et al. 1996; Yan et al. 2000).
The advantage of the $\ AB_{\nu}$ system is that the magnitude is
directly related to the physical units. The 4 Oke \& Gunn (1983)
standards are used for flux calibration in the BATC survey.  These four
stars are
BD+$17^\circ$4708, BD+$26^\circ$2606, HD84937 and HD19445. The
magnitudes of these standards were refined by several
authors. Fukugita et al. (1996) presented the latest re-calibrated
fluxes of these four standards. Their magnitudes have also been corrected
with the BATC photometric system  (Zhou et al. 2001).

When performing flat-field correction for our large-format CCD, 
a simple method was applied to reach very high quality in flat-fielding. 
This high quality in flat-fielding is achieved 
by placing an isotropic diffuser in front of the Schmidt
correction plate, and illuminating the diffuser with scattered light
from the dome screen.  Normally 12 dome flat-field images are taken
in each filter band within 24 hours of  observation.

There are two target fields in the survey for the comparison between
$z_{phot}$ and $z_{spec}$, the BATC TA03 field and
the BATC T329 field.  The TA03 field is
centered on the galaxy cluster Abell 566 with redshifts around 0.1.
The spectroscopic redshifts of the 10 central galaxies in this field 
are given by Slinglend et al. 
(1998).  The T329 field is centered on a high redshift quasar, 
located at
$\alpha=\rm {9^{h}56^{m}25\hspace{0.1cm}{.}\hspace{-0.15cm}^{s}2}$,
$\delta=+47^\circ34^{\prime}42^{\prime\prime}{\mbox{}\hspace{-0.1cm}.0}$
with $z=4.457$. The redshifts of 17 galaxies in this field 
are presented by Postman et al. (1996) and Holden et al. (1999).  These 
information can be found in NASA/Ipac Extragalactic Database (NED) at 
$http://nedwww.ipac.caltech.edu/$. 
The redshift information of the total of 27 galaxies from these two fields 
are used to check the quality of the BATC $z_{phot}$.

\section{THE DATA REDUCTION PROCEDURES}

The BATC survey images are
reduced through standard procedures, including bias subtraction, 
flat-field correction, coordinate and flux calibrations 
(see Fan et al. 1996; Zhou et al. 2001; Zhou et al. 2002 for
details).

After the basic corrections described above, 
the flat-field images and the field images observed in the same filter 
in the same night are combined, respectively.
 During combination, bad pixels and cosmic rays
are removed.  The HST Guide Star Catalog (GSC) 
(Jenkner et al. 1990) is then used for coordinate determination.  
The final RMS error in coincidence with the GSC stars is about 0.5 arcsec.
The BATC photometry code was developed based on Stetson's DAOPHOT 
procedures (Stetson 1987).  Magnitudes derived via Point Spread Function
(PSF) fitting method as well as aperture photometry method 
are given for {\bf every source detected} in the fields.  
The limiting magnitude in general 
is about $20^{m}$ with an error about 
$0\hspace{0.1cm}{.}\hspace{-0.15cm}^{m}$1 in all bands. 

{\bf PSF fitting is used basically to obtain an estimate of 
magnitude for a point source. Our PSF magnitudes were obtained 
through an automatic data reduction code, PIPELINE I, developed as a
standard procedure in the BATC multicolor sky survey (Fan et al.
1996: Zhou et al. 2001).
A distant galaxy, which is small in angular size, 
can be regarded as a point source. Although PSF fitting magnitude is 
different from the total integrated magnitude,
the shape of the SED of a galaxy should not 
change much. Furthermore, for crowded fields, 
aperture photometry may not lead to as accurate results as
PSF fitting codes. 
So PSF photometry provides an alternative approach in addition to the
aperture photometry. It will be scientifically interesting to compare 
the accuracy in redshift determination using these two methods. We have thus 
carried out the estimate of photometric redshift using the magnitudes derived
in each filter via both PSF fitting and aperture photometry.}

For larger galaxies showing obviously extended morphology,  
their magnitudes obtained via PSF fitting would have larger uncertainties 
than for smaller galaxies. In this case, the method of aperture photometry 
magnitude should be adopted.

Most galaxies in T329 field are faint and small in angular size, for which
the PSF fitting method is suitable to use.  For
comparison, we use both PSF fitting magnitudes and aperture magnitudes
to estimate redshifts of these galaxies. The results and discussions
are given in $\S$ 6. The 10 galaxies in the center of the other field, 
the Abell 566, are the 
brightest ones in this galaxy cluster 
and show obviously extended structure in the
images.  We thus use only magnitudes from aperture photometry. 
The radius of aperture adopted is 5 pixels which corresponds to a
sky projection of 8.5 arcsec.

\section{SED FITTING METHOD}

The SED fitting method is to fit the spectrum of an 
object  which should include 
several strong spectral features such
as 4000{\AA} break, Lyman-forest decrement etc. We use the
$hyperz$ program developed by Bolzonella, Miralles, \&
R\"{o}ser Pell\'{o} (2000) to estimate the redshifts of galaxies.
The standard $\chi^2$ minimization, i.e., computing and 
minimizing the deviations between photometric SED of an object
and the template SEDs obtained with the same 
photometric system, is used in the fitting process.  The
minimum $\chi^2$ indicates the best fit to the observed SED by
the set of template spectra:
\begin{equation}
\chi^{2}(z)=\sum\limits_{i=1}^{N_{filt}}\left [\frac{F_{obs,i}-b\times
 F_{temp,i}(z)} {\sigma_{i}} \right ]^{2},
\end{equation}
where $F_{obs,i}$, $F_{temp,i}$, and $\sigma_{i}$ are the observed 
fluxes, template fluxes, and the photometric uncertainty 
in filter $i$, respectively.  $b$ is the normalization constant, while
$N_{filt}$ is the number of filters used in the observations. 

In $hyperz$ program, a number of spectra templates 
can be used, including the
enlarged galaxy evolutionary library of Bruzual and Charlot (1993), as
well as the empirical template. The parameters involved in the template
construction contain SFR type, IMF, metallicity, and age of stellar
population, etc. Synthetic template that has been used the most 
is GISSEL 98 template (Bruzual \& Charlot 1993).  On the other 
hand, the empirical template generally used is
obtained through the averaged spectra of observed local field galaxies
(Coleman, Wu, \& Weedman 1980), and is suitable only for low
redshift galaxies.  The validity of direct extension to high redshift 
objects using this 
template still needs to be tested. 
It has been shown that the synthetic and empirical templates give almost
the same accuracy  for $z_{phot}$'s (Massarotti et al. 2001a).  In this
work, we used GISSEL 98 template for redshift determination.  

Fluxes given by synthetic SED models need further corrections 
for the interstellar and intergalactic medium (ISM \& IGM) extinction
effects. There are different reddening laws for the ISM extinction 
corrections.  In this paper the reddening law of
Allen (1976) for the Milky Way is adopted. The IGM, on the other
hand, affects dramatically the ionizing continuum blueward of 
redshifted $Ly\alpha$, which makes Lyman-forest the most
important spectral feature for objects with redshifts beyond
2.0.  However, due to the survey depth of the BATC images, 
almost all the objects observed have redshifts less than 0.5, for which  
the Lyman-forest has minimal effect and 
does not enter into the wavelength range of concern and
thus has no effect on our analysis. 
We thus do not take into account the extinction effect 
of IGM. 

The most obvious and useful spectral feature in 
redshift determination with the BATC system is then 
the 4000{\AA} 
Balmer break, which falls in the redshifted wavelength
range of 4000 to 6000{\AA} approximately, 
corresponding to the BATC filters from
$b$ to $h$.  The observations made with 
filters whose central wavelengths are shorter than 
6000{\AA}, are therefore extremely crucial for the success of 
this project.  We will reinforce this point in $\S$ 4.

\section{SIMULATION TEST OF BATC PHOTOMETRIC REDSHIFT}

\subsection{Comparison between BATC and  
$\ UBVRI$ Filter Systems}

In the $hyperz$ program, the procedure $make\_catalog$ checks
the self-consistency of SED fitting method for a given photometric
system. To examine the dependence of $z_{phot}$ uncertainty on
photometric errors, we use this program to build a catalog containing
simulated galaxies of different redshifts and types.  
Gaussian distribution of magnitude error in different filters is
assumed.  For the comparison between the BATC filter system, and 
the $\ UBVRI$ filter system of Canada-France-Hawaii Telescope (CFHT), 
we created a catalog of 1000 galaxies using the total of 20 filters (15
BATC and 5 $\ UBVRI$ filters) simultaneously.  It is thus 
guaranteed that the comparison between the 
two systems is done for the same sample with 
the same redshifts. The redshift range in the simulation is set to be $z=0$ to
6.  $z_{phot}$s of the 1000 galaxies are then estimated using the BATC
system and $\ UBVRI$ system, respectively. Photometric
uncertainties of $0\hspace{0.1cm}{.}\hspace{-0.15cm}^{m}$03,
$0\hspace{0.1cm}{.}\hspace{-0.15cm}^{m}$05,
$0\hspace{0.1cm}{.}\hspace{-0.15cm}^{m}$1,
$0\hspace{0.1cm}{.}\hspace{-0.15cm}^{m}$2 and
$0\hspace{0.1cm}{.}\hspace{-0.15cm}^{m}$3  are assumed. 
To maximize the efficiency of computing when fitting,  we choose the 
increment in redshift to be 
$z_{step}=0.05$, and  $\ Av_{step}=0.2$ in $\ Av$  range of 0 to 1.2
following the values given by Bolzonella et al. (2000).

Figure 2 shows the results of photometric and catalog redshifts
($z_{phot}$ vs. $z_{cat}$) with uncertainties of
$0\hspace{0.1cm}{.}\hspace{-0.15cm}^{m}$05,
$0\hspace{0.1cm}{.}\hspace{-0.15cm}^{m}$10 and
$0\hspace{0.1cm}{.}\hspace{-0.15cm}^{m}$20, respectively. 
The quality of the 
$z_{phot}$ estimation with the two systems are summarized in Table 2.

\begin{figure}
\figurenum{2} \epsscale{0.75} \plotone{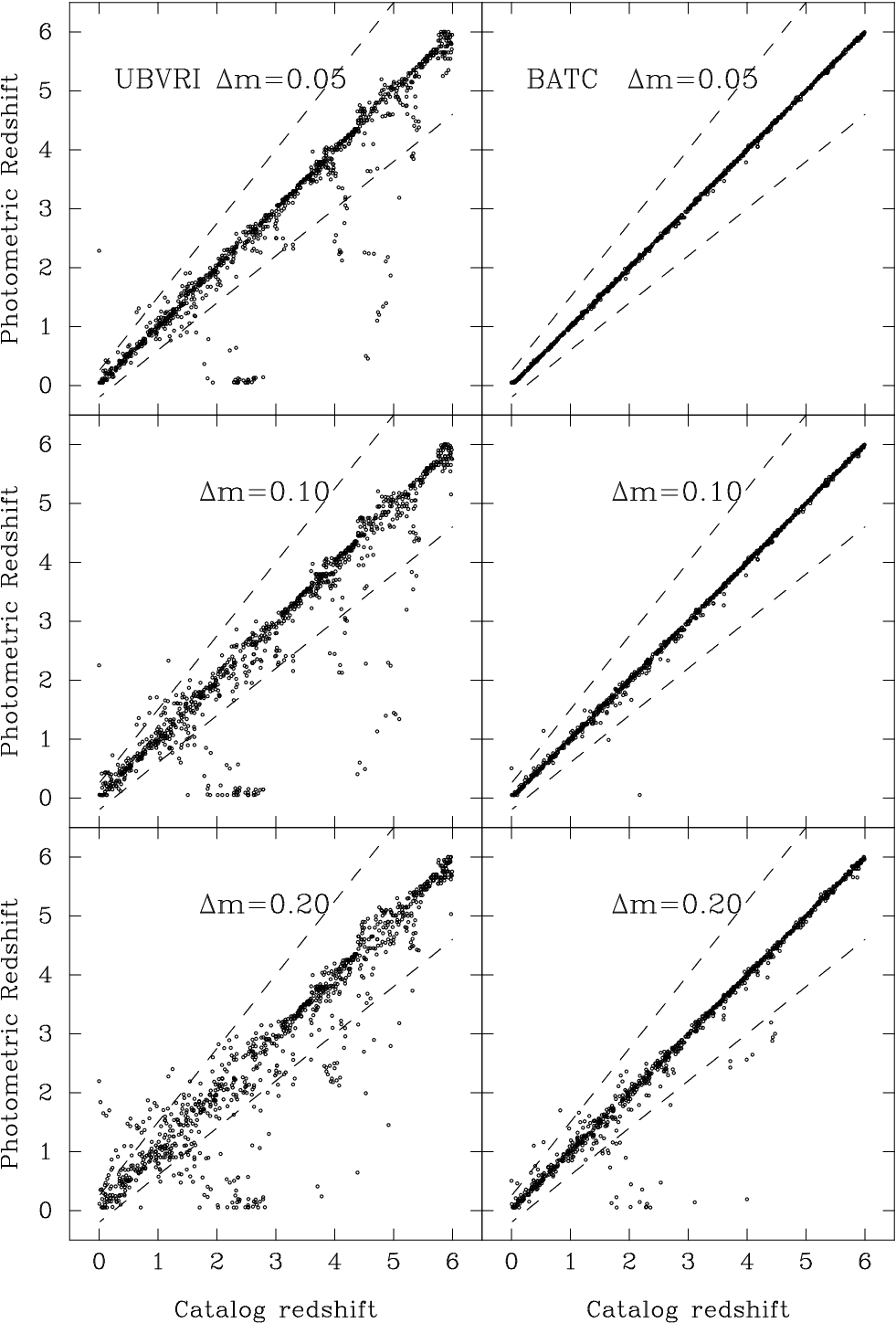}
\caption{Plot of the catalog redshifts versus the fitted redshifts 
($z_{cat}$ vs. $z_{phot}$) with the BATC system and the $\ UBVRI$ 
system with photometric errors of 
$0\hspace{0.1cm}{.}\hspace{-0.15cm}^{m}05$,
$0\hspace{0.1cm}{.}\hspace{-0.15cm}^{m}1$,
$0\hspace{0.1cm}{.}\hspace{-0.15cm}^{m}2$, respectively. Redshift
range is from $z=0$ to 6. Dashed lines separates the 
catastrophic failures from the reasonable fits. The circles located 
between the two dashed lines are regarded as good estimations.}
\end{figure}

The quality of $z_{phot}$ estimation with simulated
photometric errors is evaluated using the following parameters:
$\ l$, $\overline {\Delta z}$, and $\sigma_{z}$.  
 
The first one, $\ l$, is the catastrophic percentage of the determination, 
which is the
ratio of incorrect determinations over the total number of estimations, 
\begin{equation}
\ l=\frac {N_{incrt}} {N_{total}},
\end{equation}
$N_{incrt}$ is the number obtained using the following formula,
\begin{equation}
\frac {1}{threshold} < \frac {1+z_{phot}} {1+z_{spec}}
< {threshold},
\end{equation}
here the threshold is taken to be 1.25, which means that the 
difference between 
the estimated SED and original SED is greater than -0.20 and less than 
0.25 at a given wavelength.
The systematic error, $\overline {\Delta z}$, is defined as the mean
difference  $\overline {\Delta z}={\sum {\Delta z}}/{N_g}$.
The standard deviation of the estimation excluding the 
catastrophic identifications, $\sigma_{z}$, is given by, 
\begin{equation}
\sigma_{z}^{2}=\sum\limits_{i=1}^{N_g} \frac {(\Delta z
-\overline{\Delta z})^{2}}{N_g-1},
\end{equation}
where $N_g$ is the number of galaxies excluding $\ l$.

We first discuss the results of estimation of $z_{phot}$ using the 
BATC system. From
Table 2, we can see that, for the smallest photometric uncertainty
$\Delta m=0\hspace{0.1cm}{.}\hspace{-0.15cm}^{m}03$, we obtain
the best fit with $\sigma_{z}=0.019$ and $\ l=0$, which means that all the
1000 galaxies are estimated correctly.  The choice of the small
photometric errors anywhere between 0.03 and 0.05 does not affect
the results significantly. With the increase of photometric
uncertainties from $\Delta m=0\hspace{0.1cm}{.}\hspace{-0.15cm}^{m}05$
to $0\hspace{0.1cm}{.}\hspace{-0.15cm}^{m}1$, $\sigma_z$ and $\ l$
also increase, i.e., from $\sigma_{z}=0.021$ and $\ l=0$ to 0.042 and
$0.4\%$. Figure 2 shows this trend, especially when 
$\Delta m=0\hspace{0.1cm}{.}\hspace{-0.15cm}^{m}2$ the scatter becomes
significantly larger. The increasing scatter for 
$\Delta m=0\hspace{0.1cm}{.}\hspace{-0.15cm}^{m}2$ is caused by the
ambiguity in the spectra created with large 
photometric uncertainty which leads to confusion when the program
tries to identify certain features.  Therefore, as long as the accuracy of
our photometry meets the criterion, reasonable
redshift estimation is guaranteed.   

With Table 2 and Figure 2, we see the distinct advantage of the
BATC photometric system over the $\ UBVRI$ system. For
$\Delta m=0\hspace{0.1cm}{.}\hspace{-0.15cm}^{m}05$, the performance
is $\sigma_z=0.021$, $\ l=0$ for the BATC system and $\sigma_z=0.174$,
$\ l=6.7\%$ for the $\ UBVRI$ system.  At this level of uncertainty, a large number
of galaxies have already dropped out of the acceptable region for 
the $\ UBVRI$ system.  On the other hand, the deviation is very small 
with the BATC system at this level of uncertainty, 
and all estimates are within the acceptable range. 

Figure 2 also shows large dispersion and persistent scatter 
even for the smallest photometric uncertainty using the $\ UBVRI$
system.  This is due to the fewer number and larger bandwidth of filters, 
which makes the system less sensitive to the delicate
spectral features.  For example, in redshift range between
2.0 and 3.0, the $\ UBVRI$ system is less sensitive to the difference
between redshifted Lyman-forest and rest-frame 
Balmer break.   

From Table 2, we can also see that the performance of the BATC
system with even the largest observational uncertainty 
$\Delta m=0\hspace{0.1cm}{.}\hspace{-0.15cm}^{m}2$, is still better than 
the $\ UBVRI$ system with the smallest uncertainty 
$\Delta m=0\hspace{0.1cm}{.}\hspace{-0.15cm}^{m}03$.

A number of studies have used $\ UBVRI$ system for redshift determination. 
These studies with the broad-band filters 
give a general accuracy of 
around $\sigma_{z}\sim 0.15$ (Fontana et al. 2000, 
Massarotti et al. 2001a \& b, Le Borgne \& Rocca-Volmerange 2002). The
relevant studies and results are summarized in Table 3.

The study in this paper shows that the accuracy of 
redshift determination with the BATC system can 
reach $0.02-0.03$ with  $\Delta m$ from
$0\hspace{0.1cm}{.}\hspace{-0.15cm}^{m}05$ to
$0\hspace{0.1cm}{.}\hspace{-0.15cm}^{m}1$.  This conclusion is in
good agreement with that from Hickson et al. (1994), who has 
performed a computer simulation for multinarrowband system of 40 
bands to investigate the potential of determining galaxy morphological 
type and redshift of galaxy. The results in their study show that, 
for a signal-to-noise ratio 
of 10, $\sigma_{z}$ is less than 0.02; for a signal-to-noise ratio of 3, 
$\sigma_{z}$ is 0.06 with redshift $z<0.5$ and about 0.03 with $0.5<z<1.0$.
Thus the accuracy of 
our simulated photometric redshift with the BATC 15-color system 
agrees very well with Hickson et al. (1994).

However, the limiting magnitude of the BATC system 
is about $20\hspace{0.1cm}{.}\hspace{-0.15cm}^{m}0$. 
At this level of brightness, only objects with redshift
less than 0.5 can be observed, plus a few luminous 
high-redshift quasars, using this system.  
In order to test how well the system performs in the study
of the structure of local universe, we repeat the simulation with a 
redshift range of $z=0-0.5$. The $z_{step}$ is refined to 0.005 in order to
carry out better differentiation.  The results with $z_{step}$
of 0.05 are given in Table 4, and the plot of $z_{cat}$ vs. $z_{phot}$
is shown in Figure 3.

\begin{figure}
\figurenum{3} \epsscale{0.75} \plotone{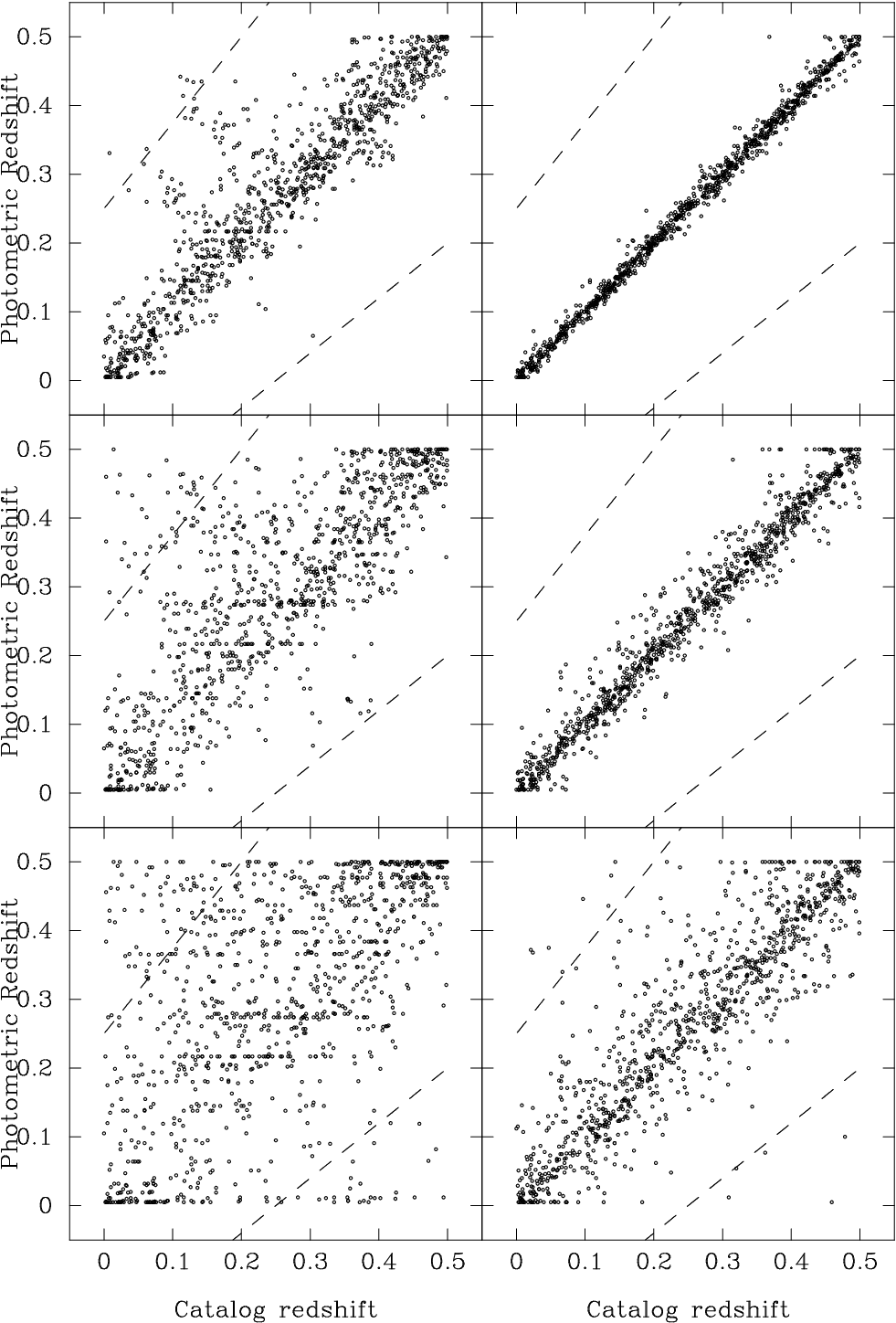}
\caption[]{Plot of catalog redshifts versus fitted redshifts
($z_{cat}$ vs. $z_{phot}$) as Figure 2 but with the $z_{step}$
refined to be 0.005.  The fitted
redshift range is from 0 to 0.5.}
\label{fig3}
\end{figure}

From Table 4 we see that, for the BATC system, 
the $\sigma_{z}$ improves when $z_{step}$ 
drops from 0.05 to 0.005.  But the improvement gradually diminishes
when the magnitude error becomes larger.  For $\ UBVRI$ system, 
however, the $\sigma_{z}$ becomes larger with increased magnitude
errors, when $z_{step}$ is refined to 0.005.  As to the 
$\overline{\Delta z}$, it improves by about the same amount for
both systems when $z_{step}$ becomes 0.005.  This is partly due
to the fact that the accuracy of $z_{phot}$ for lower redshift objects 
is better than that for higher redshift ones (Bolzonella et
al. 2000; Massarotti et al. 2001b).  The refinement of 
$z_{step}$ also allows more flexibility when performing the fitting.  

From this test it is clear that, to estimate $z_{phot}$ precisely, we
should not only adopt smaller $z_{step}$ for fitting but also process the 
photometry as accurately as possible.  From the comparison and analysis
above, the BATC multicolor photometric system reveals its distinct
advantage on $z_{phot}$ estimation, especially for the low redshift
objects.  The typical value of $\sigma_{z}$ can be as low as  
$0.02\sim 0.03$.

It should be pointed out that since the simulated catalog is created from
templates, there exists the problem of incompleteness.  Because 
all the tests are performed using these
templates, when it comes to the analysis of real observations, 
certain spectra will not
find their counterparts in the catalog, which consequently degrades the
overall fitting quality.  

The data in Tables 2 and 4 also indicate that, in redshift range from 0
to 0.5, the systematic errors are all positive, which means that in
this range the redshifts tend to be overestimated. On the
other hand, for $z$'s from 0 to 6, the systematic errors are all negative
which means that the redshifts are being underestimated. This 
result agrees with the findings by Massarotti et al. (2001b).

\subsection{Optimization of Redshift Survey from Filter Sets}

Balmer Jump is the dominant spectral feature in wavelengths 
shorter than 6000{\AA} for galaxies with redshifts from 0 to 0.5.  
Below we consider if we could only use several crucial filters of
shorter wavelengths to
achieve the same goal, estimating $z_{phot}$ properly,
but at the same time maximize the 
observational efficiency.  
We perform this test by deleting longer wavelength filters, one at 
a time, from
$p$ to $h$ (wavelength coverage decreases from 10000 to 6000{\AA}). The
photometric error is chosen to be the typical value 
$\Delta m=0\hspace{0.1cm}{.}\hspace{-0.15cm}^{m}05$. 
The redshift range is set from
$z=0$ to 0.5 with $z_{step}=0.005$.  The results are summarized in
Table 5.

In Table 5, column 1 lists the number of filters that are used when
performing the estimation.  Column 2 lists the corresponding labels of filters
used.  $\sigma_z$ and $\ l$ are defined as in Table 4.  The
first line in this table is the result using all 15 filters.  No obvious
degradation is seen until in the last case, which indicates that we can 
obtain accurate $z_{phot}$ estimation  
for low redshift objects using the BATC system with only
8 filters, from $a$ to $h$.

\section{COMPARISON BETWEEN PHOTOMETRIC AND SPECTROSCOPIC REDSHIFTS}

We have generated a set of 15-color SEDs for 27 galaxies 
with known $z_{spec}$s, which can be used to check the accuracy of
$z_{phot}$ obtained with the BATC system.

We first estimate the $z_{phot}$'s for the 10 central member galaxies of
Abell 566. Since these galaxies are the brightest in the cluster, their
SEDs can be obtained with small photometric uncertainties. We 
adopt a value of 
$0\hspace{0.1cm}{.}\hspace{-0.15cm}^{m}05$ for 
the photometric uncertainty, which includes 
errors from observation and from subsequent flux calibration. 
By applying $hyperz$ to the spectra with 
parameters of $z=0-0.5$, 
$z_{step}=0.05$, $\ Av$ as $0-0.3$, and $\ Av_{step}=0.03$, 
$z_{phot}$'s for the 10 galaxies are obtained. The results are listed
in Table 6. Here the range of $\ Av$ is inferred from
the best fit.  We then refine the redshift step to 
$z_{step}=0.005$.  The results with the refined 
step are given in column 7 and 8 in Table 6.  
The accuracy remains about the same for the two choices of 
$z_{step}$. However, the 
systematic error is apparently improved from $\overline {\Delta z}=-0.007$ to
$-0.002$.  We thus confirm that our estimation can be improved by using 
smaller fitting steps, and that the determination of 
photometric redshift can reach a higher 
precision for bright galaxies with smaller photometric errors.

Secondly, we also obtain the $z_{phot}$'s for the galaxies in T329 field 
using aperture and PSF photometry, respectively.
The parameters used are the same as above except for the photometric
errors. For all of the 17 galaxies, the dispersion of the measurements 
$\sigma_{z}$ is 0.021 for PSF photometry and 0.055 for aperture
photometry.  For the majority of these galaxies, the photometric 
redshifts are almost the same.  For the other several galaxies, 
the relatively large deviations 
of $z_{phot}$ are due to the differences in SED shapes. {\bf In addition, 
we examined the images carefully and found that the majority of sources 
with relatively large deviations using aperture photometry are objects 
with other objects nearby. Therefore it is apparent that PSF fitting 
is superior over the method of aperture photometry, especially when dealing
with crowded fields.}  The comparison of 
$z_{spec}$s to $z_{phot}$'s derived using two different methods are 
shown in Table 7. And $z_{spec}$ vs. $z_{phot}$ are plotted for 
both galaxies in fields of TA03 and T329 in Figure 4.
$z_{phot}$s of TA03 field are obtained using aperture photometry and those of 
T329 field are obtained using PSF photometry.
 
\begin{figure}
\figurenum{4} 
\plotone{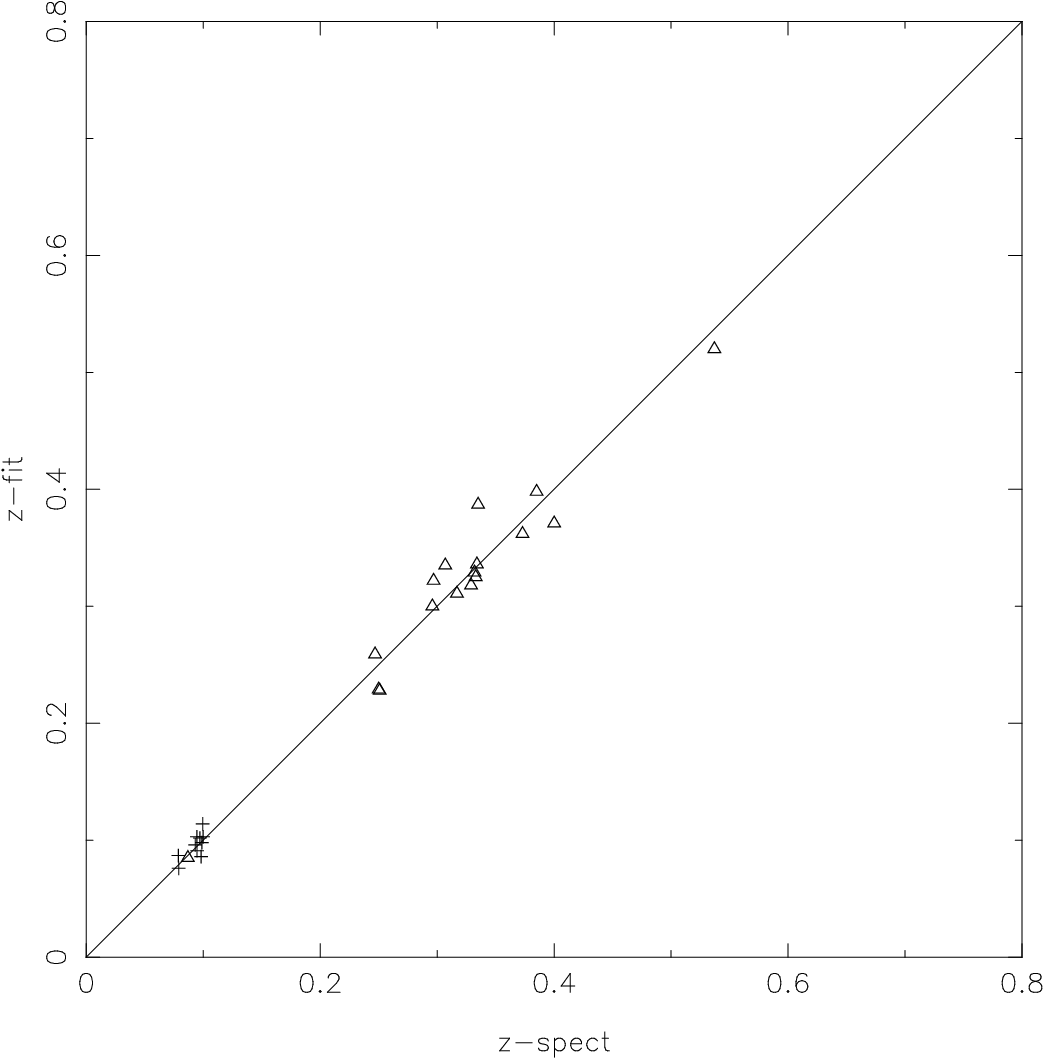}
\caption{Plot of spectroscopic redshifts versus fitted redshifts ($z_{spec}$
vs. $z_{phot}$) of the galaxies in fields of TA03 and T329. Triangles mark 
sources in field of T329 and crosses represent sources in field of TA01.
$z_{phot}$s of TA03 field are obtained using aperture photometry and those of 
T329 field are obtained using PSF photometry.}

\end{figure}

\section{SUMMARY}
In this paper, with the help of $hyperz$, 
we examine the accuracy of redshift estimation  
by comparing the BATC 15 intermediate-band photometric system 
to the $\ UBVRI$ broad-band photometric system using simulated
spectra.  We find that with the BATC system we can obtain
fairly accurate redshift estimation.  This advantage comes from the 
careful selection of the 15-color filter set in the begginning of the
BATC survey.  
The $z_{phot}$ determination of the
spectroscopic sample in the BATC fields is also checked. The main
results are listed as follows:

1. The uncertainty in photometric redshifts comes mainly 
from the
photometric errors.   We have made assessment of the accuracy 
with simulation. The dispersion can reach as low as 
$\sigma_{z}=0.02\sim 0.03$ with
almost no catastrophic dropout for the typical photometric
uncertainty from  $\Delta m=0\hspace{0.1cm}{.}\hspace{-0.15cm}^{m}05$
to  $\Delta m=0\hspace{0.1cm}{.}\hspace{-0.15cm}^{m}1$;

2. The objects that can be observed with BATC survey are 
generally limited to redshift range of 0 to 0.5,  hence the filters
whose central wavelengths are shorter than
6000{\AA} are especially important for the detection of
the 4000{\AA} Balmer break.  It has further been shown that
we can use only the filters blueward than 6000{\AA} for the 
accurate determination of redshift and save significant amount
of telescope time;

3. For the 10 brightest galaxies centered in
Abell 566, the results show that the accuracy of photometric 
redshift determination is
$\sigma_{z}=0.008$ for $z_{step}$ of 0.05 and 0.005, with 
systematic errors of $\overline {\Delta z}=-0.007$ and $-0.002$, respectively. 
For the 17 galaxies, which have spectroscopic measurements in NED, the
accuracy is $\sigma_{z}=0.021$.

\acknowledgments
We would like to thank the anonymous referee for his/her
insightful comments and suggestions which improved this paper
significantly.
The BATC Survey is supported by the Chinese Academy of Sciences, 
the Chinese National Natural Science Foundation and 
the Chinese State Committee of Sciences and Technology.
The project is also supported in part by the National Science Council 
in Taiwan under the grant NSC 89-2112-M-008-021. 
The project is also supported in part 
by the National Science Foundation (grant INT 93-01805) and
by Arizona State University, the University of Arizona and Western
Connecticut State University.

\clearpage
\setcounter{table}{0}
\begin{table}[ht]
\caption[]{The central wavelengths and effective bandpasses of
 the 15  BATC filters.\\}
\vspace{0.3cm}
\begin{tabular}{cccc}
\hline \hline 
No. & Filter & Wavelength ({\AA}) & bandpass ({\AA}) \\
\hline
1 & $a$ & 3372.1 & 337.85 \\
2 & $b$ & 3895.3 & 266.65 \\
3 & $c$ & 4202.4 & 282.07 \\
4 & $d$ & 4547.4 & 355.53 \\
5 & $e$ & 4873.3 & 347.12 \\
6 & $f$ & 5248.4 & 331.49 \\
7 & $g$ & 5784.7 & 271.67 \\
8 & $h$ & 6074.3 & 289.77 \\
9 & $i$ & 6710.8 & 497.00 \\
10 & $j$ & 7011.3 & 170.62 \\
11 & $k$ & 7527.5 & 191.91 \\
12 & $m$ & 8025.4 & 260.27 \\
13 & $n$ & 8518.2 & 185.40 \\
14 & $o$ & 9173.8 & 269.48 \\
15 & $p$ & 9724.7 & 278.20 \\
\hline
\end{tabular}
\end{table}

\clearpage
\setcounter{table}{1}
\begin{table}[ht]
\caption[]{Comparison of $z_{phot}$ between the two
systems. $\sigma_{z}$ is the dispersion excluding {\bf those catastrophic 
failures, $\  l$ is the fraction of galaxies with redshift errors 
greater than 0.25 in $1+z$.} 
$\overline{\Delta z}$ is the mean difference excluding $l$.
Redshift range is from 0 to 6.}
\vspace{0.5cm}
\begin{tabular}{c|cc|cc|cc}
\hline
\hline
$\Delta m$  & \multicolumn{2}{|c|}{$\sigma_{z}$} & \multicolumn{2}{c|}
{$l$(\%)} & \multicolumn{2}{c}{$\overline {\Delta z}$}\\ \hline
  & BATC & $\ UBVRI$ & BATC & $\ UBVRI$ & BATC & $\ UBVRI$ \\ \hline
0.03 & 0.019 & 0.163 & 0.0 & 4.3  & $-0.003$  & $-0.037$ \\
0.05 & 0.021 & 0.174 & 0.0 & 6.7  & $-0.003$  & $-0.043$ \\
0.10 & 0.042 & 0.195 & 0.4 & 8.9  & $-0.007$  & $-0.046$ \\
0.20 & 0.084 & 0.234 & 3.8 & 13.8 & $-0.014$  & $-0.056$ \\
0.30 & 0.133 & 0.324 & 8.0 & 20.7 & $-0.029$  & $-0.050$ \\ \hline
\end{tabular}
\end{table}

\clearpage
\setcounter{table}{2}
\begin{table}[ht]
\caption[]{Summarized results of some other authors.}
\vspace{0.5cm}
\begin{tabular}{lllllll}
\hline
\hline
 & sample & filters & redshift range & $\Delta m$ & 
$\sigma_{z}$ & $\overline{\Delta z}$ \\ \hline
Bolzonella et al. (2000) & model & $UBVRI$   & $z<0.4$ & 0.05 & 0.07 & 0.03   \\
                         &       &         &         & 0.10 & 0.09 & 0.03   \\
                         &       &         &         & 0.20 & 0.20 & 0.11   \\
                         &       &         &         & 0.30 & 0.28 & 0.20   \\
Fontana    et al. (2000) & real  & $UBVRIJK$ & $z<1.5$ &     & 0.08 &        \\
                         &       &         & $z>2.0$ &     & 0.32 & $-0.144$  \\
Massarotti et al. (2001) & real  & $UBVIJHK$ & $z<1.5$ &     & 0.070 & $-0.001$   \\
                         &       &         & $z>2.0$ &     & 0.177 & $-0.156$   \\
Fern\'{a}ndez-Soto et al. (2001) & real & $UBVIJHK$ & $z<1.5$     &     & 0.110 & 0.002  \\
                                 &      &         & $2.0<z<4.0$ &     & 0.285 & 0.06  \\
Borgne et al. (2002)     & real  & $UBVI$           & $z<1.5$ &     & 0.318 & $-0.127$  \\
                         &       & $UBVIJHK$        & $z<1.5$ &     & 0.098 & 0.021 \\
Wolf et al. (2001)       & real  & $16-$ color       &        &     & 0.03  & 0    \\
Hickson et al. (1994)    & model & 40 bands    & $z<0.5$ & s/n=10 & $<0.02$ &   \\
                         &       & (simulation)&         & s/n=3  & 0.06 &   \\
                         &       &             & $0.5<z<1.0$  & s/n=10   & $<0.01$ &  \\
                         &       &             &              & s/n=3    & 0.03  &  \\       
\hline
\hline
\end{tabular}
\end{table}

\clearpage
\setcounter{table}{3}
\begin{table}[ht]
\caption[]{The dispersion comparison of different $z_{step}$ 0.05 with 0.005. 
Redshift range from $z=0$ to 0.5.}
\vspace{0.5cm}
\begin{tabular}{c|c|cc|cc|cc}
\hline
\hline
$\Delta m$ & $z_{step}$ & \multicolumn{2}{|c|}{$\sigma_{z}$} 
& \multicolumn{2}{|c|}{$\ l$} & \multicolumn{2}{c}{$\overline{\Delta z}$} \\ 
\hline
  &  & BATC  & $\ UBVRI$ & BATC & $\ UBVRI$  & BATC & $\ UBVRI$ \\ \hline
0.03 & 0.05  & 0.013 & 0.040 & 0.000 & 0.001 & 0.005 & 0.016 \\
     & 0.005 & 0.006 & 0.039 & 0.000 & 0.000 & 0.000 & 0.014 \\
0.05 & 0.05  & 0.016 & 0.055 & 0.000 & 0.010 & 0.005 & 0.024 \\
     & 0.005 & 0.012 & 0.055 & 0.000 & 0.007 & 0.001 & 0.021 \\
0.10 & 0.05  & 0.029 & 0.079 & 0.002 & 0.033 & 0.008 & 0.033 \\
     & 0.005 & 0.028 & 0.081 & 0.000 & 0.034 & 0.005 & 0.029 \\
0.20 & 0.05  & 0.062 & 0.103 & 0.023 & 0.095 & 0.016 & 0.042 \\
     & 0.005 & 0.062 & 0.106 & 0.013 & 0.096 & 0.012 & 0.037 \\
0.30 & 0.05  & 0.086 & 0.116 & 0.042 & 0.129 & 0.017 & 0.040 \\
     & 0.005 & 0.086 & 0.120 & 0.041 & 0.130 & 0.012 & 0.034 \\ \hline
\end{tabular}
\end{table}

\clearpage
\setcounter{table}{4}
\begin{table}[ht]
\caption[]{The estimated results using different selected filter sets in 15.}
\vspace{0.5cm}
\begin{tabular}{ccccc}
\hline \hline No. & I.D. & $\sigma_z$ & $\ l$ & $\overline{\Delta z}$ \\ \hline
15 &    $\ a-p$  &  0.011  &  0.000 &  0.001 \\
14 &    $\ a-o$  &  0.011  &  0.000 &  0.001 \\
13 &    $\ a-n$  &  0.012  &  0.000 &  0.001 \\
12 &    $\ a-m$  &  0.012  &  0.000 &  0.001 \\
11 &    $\ a-k$  &  0.013  &  0.000 &  0.001 \\
10 &    $\ a-j$  &  0.013  &  0.000 &  0.001 \\
9  &    $\ a-i$  &  0.014  &  0.000 &  0.001 \\
8  &    $\ a-h$  &  0.016  &  0.000 &  0.002 \\
7  &    $\ a-g$  &  0.018  &  0.004 &  0.002 \\
\hline
\end{tabular}
\end{table}

\clearpage
\setcounter{table}{5}
\begin{table}[ht]
\caption[]{The fitting results of BATC TA03 field (Abell 566).}
\vspace{0.5cm}
\begin{tabular}{cccc|cc|cc}
\hline
\hline
 & & & & \multicolumn{2}{|c}{for $z_{step}=0.05$} & 
 \multicolumn{2}{|c}{for $z_{step}=0.005$} \\ \hline
No. & $\alpha$(J2000) & $\delta$(J2000) & $z_{spec}$ & $z_{phot}$&
 $\Delta z$ & $z_{phot}$ & $\Delta z$ \\ \hline
1 & 07:04:43.12 & 63:18:38.9 & 0.09829  &  0.0950  &    0.003   & 0.0860 &   0.012   \\
2 & 07:06:04.05 & 63:12:39.5 & 0.07884  &  0.0900  &  $-0.011$  & 0.0870 & $-0.008$ \\
3 & 07:04:07.90 & 63:08:06.7 & 0.09725  &  0.1000  &  $-0.003$  & 0.1020 & $-0.005$ \\
4 & 07:04:28.86 & 63:18:38.0 & 0.09479  &  0.0950  &    0.000   & 0.0910 &   0.004  \\
5 & 07:04:39.85 & 63:19:18.3 & 0.09881  &  0.1000  &  $-0.001$  & 0.0980 &   0.001  \\
6 & 07:05:33.97 & 63:15:26.4 & 0.10007  &  0.1050  &  $-0.005$  & 0.1030 & $-0.003$ \\
7 & 07:06:17.92 & 63:06:50.1 & 0.07910  &  0.1050  &  $-0.026$  & 0.0760 &   0.003   \\
8 & 07:03:29.96 & 63:15:16.7 & 0.09969  &  0.1100  &  $-0.010$  & 0.1140 & $-0.014$ \\
9 & 07:03:46.65 & 63:19:27.0 & 0.09463  &  0.1050  &  $-0.010$  & 0.1030 & $-0.008$ \\
10 & 07:05:33.88 & 63:05:24.2 & 0.09319 &  0.0950  &  $-0.002$  & 0.0960 & $-0.003$ \\
\hline
\multicolumn{7}{l}{for $z_{step}=0.05$: $\overline{\Delta z}=-0.007$,  
$\sigma_{z}=0.008$}\\  
\multicolumn{7}{l}{for $z_{step}=0.005$: $\overline{\Delta z}=-0.002$, 
$\sigma_{z}=0.008$} \\ \hline
\end{tabular}
\end{table}

\clearpage
\setcounter{table}{6}
\begin{table}[ht]
\caption[]{The photometric and  spectroscopic  redshifts of
 galaxies in BATC T329 field.}
\vspace{0.5cm}
\begin{tabular}{cccc|cc|cc}
\hline
\hline
 & & & & \multicolumn{2}{|c}{PSF photometry} & 
\multicolumn{2}{|c}{aperture photometry} \\ \hline
 No. & $\alpha$(J2000)  & $\delta$(J2000)  & $z_{spec}$ &
$z_{phot}$ & $\Delta z$ & $z_{phot}$ & $\Delta z$\\ \hline
  1 & 09:54:38.24 & 47:10:26.2 & 0.251 & 0.228 &  0.023    &   0.237 &   0.014    \\
  2 & 09:54:39.00 & 47:15:48.4 & 0.400 & 0.371 &  0.029    &   0.360 &   0.040    \\
  3 & 09:55:03.45 & 47:28:34.3 & 0.329 & 0.318 &  0.011    &   0.326 &   0.003    \\
  4 & 09:55:06.08 & 47:29:05.2 & 0.334 & 0.336 & $-0.002$  &   0.479 &  $-0.145$  \\
  5 & 09:55:08.56 & 47:29:43.3 & 0.333 & 0.325 &  0.008    &   0.323 &   0.010    \\
  6 & 09:55:04.20 & 47:29:50.4 & 0.385 & 0.398 & $-0.013$  &   0.311 &   0.074    \\
  7 & 09:55:08.96 & 47:29:54.0 & 0.332 & 0.329 &  0.003    &   0.330 &   0.002    \\
  8 & 09:55:12.78 & 47:30:32.1 & 0.335 & 0.387 & $-0.052$  &   0.415 &  $-0.080$  \\
  9 & 09:54:03.77 & 47:40:04.5 & 0.247 & 0.259 & $-0.012$  &   0.273 &  $-0.026$  \\
  10 & 09:54:05.22 & 47:41:32.5 & 0.250 & 0.229 &  0.021   &   0.237 &   0.013    \\
  11 & 09:54:00.67 & 47:58:05.1 & 0.537 & 0.520 &  0.017   &   0.518 &   0.019    \\
  12 & 09:54:24.46 & 47:58:41.1 & 0.307 & 0.335 & $-0.028$ &   0.346 &  $-0.039$  \\
  13 & 09:54:26.93 & 47:58:53.9 & 0.296 & 0.300 & $-0.004$ &   0.279 &   0.017    \\
  14 & 09:54:28.62 & 47:58:57.3 & 0.317 & 0.311 &  0.006   &   0.231 &   0.086    \\
  15 & 09:54:24.43 & 47:58:58.9 & 0.297 & 0.322 & $-0.025$ &   0.350 &  $-0.053$  \\
  16 & 09:54:30.62 & 48:00:21.0 & 0.373 & 0.362 &  0.011   &   0.339 &   0.034    \\
  17 & 09:53:51.81 & 47:55:56.1 & 0.087 & 0.085 &  0.002   &   0.084 &   0.003    \\ \hline
 \multicolumn{6}{l}{for PSF photometry: $\overline{\Delta z}=0.000$,  $\sigma_{z}=0.021$} \\ 
 \multicolumn{6}{l}{for aperture photometry: $\overline{\Delta z}=-0.002$,  $\sigma_{z}=0.055$} \\ \hline
\end{tabular}
\end{table}

\end{document}